\documentclass{article}

\usepackage{arxiv}

\usepackage[utf8]{inputenc} 
\usepackage[T1]{fontenc}    
\usepackage{hyperref}       
\usepackage{url}            
\usepackage{booktabs}       
\usepackage{amsfonts}       
\usepackage{microtype}      
\usepackage{lipsum}
\usepackage{graphicx}
\graphicspath{ {./images/} }

\usepackage{subcaption} 
\usepackage{xspace}
\usepackage{amsmath}
\usepackage{siunitx}

\title{Molecular Machine Learning in\\ Chemical Process Design}

\author{
  Jan G. Rittig \textsuperscript{1,2}, \quad Manuel Dahmen \textsuperscript{3}, \quad Martin Grohe \textsuperscript{4}, \quad Philippe Schwaller \textsuperscript{2,5}, \quad Alexander Mitsos \textsuperscript{1,6,3,*} \vspace*{2mm}\\
	\textsuperscript{1}{Process Systems Engineering (AVT.SVT), RWTH Aachen University} \\
	\textsuperscript{2}{Laboratory of Artificial Chemical Intelligence (LIAC), Institute of Chemical Sciences and Engineering, EPFL}\\
	\textsuperscript{3}{Institute of Climate and Energy Systems ICE-1: Energy Systems Engineering, Forschungszentrum Jülich GmbH}\\
	\textsuperscript{4}{Lehrstuhl Informatik 7, RWTH Aachen University}\\
	\textsuperscript{5}{National Centre of Competence in Research (NCCR) Catalysis, EPFL}\\
	\textsuperscript{6}{JARA Center for Simulation and Data Science (CSD)}\\
	\textsuperscript{*}{Corresponding author, \texttt{amitsos@alum.mit.edu}}\\
}

\begin{document}

\maketitle
\begin{abstract}

We present a perspective on molecular machine learning (ML) in the field of chemical process engineering.
Recently, molecular ML has demonstrated great potential in (i) providing highly accurate predictions for properties of pure components and their mixtures, and (ii) exploring the chemical space for new molecular structures.  
We review current state-of-the-art molecular ML models and discuss research directions that promise further advancements. 
This includes ML methods, such as graph neural networks and transformers, which can be further advanced through the incorporation of physicochemical knowledge in a hybrid or physics-informed fashion.
Then, we consider leveraging molecular ML at the chemical process scale, which is highly desirable yet rather unexplored.
We discuss how molecular ML can be integrated into process design and optimization formulations, promising to accelerate the identification of novel molecules and processes.
To this end, it will be essential to create molecule and process design benchmarks and practically validate proposed candidates, possibly in collaboration with the chemical industry.

\end{abstract}

\section{Introduction}\label{sec:Intro}

\noindent Machine learning (ML) has advanced molecular property prediction and design.
Over the last years, a variety of ML methods, such as graph neural networks (GNNs)~\cite{Heid.2024, Reiser.2022, SanchezMedina.2023, Schweidtmann_FuelGNN.2020, Vermeire.2021}, transformers~\cite{Winter.2022}, and matrix completion methods (MCMs)~\cite{Jirasek.2020}, have been extensively applied and further developed for predicting properties of molecules and their mixtures. 
These ML methods have achieved high prediction accuracies, outperforming well-established methods in the field of chemical engineering~(ChemE) such as the group contribution method UNIFAC~\cite{Fredenslund.1975} and the quantum mechanics- and statistical thermodynamics-based model COSMO-RS~\cite{Klamt.2010}. 
Coupling ML with physicochemical knowledge can further greatly enhance or even ensure thermodynamic consistency of the predictions and decrease data required for training~\cite{rittig2024thermodynamics, specht2024hanna}.
Moreover, generative ML models have emerged for computer-aided molecular design~(CAMD)~\cite{Bilodeau2022, Elton2019, du2024machine}, providing new possibilities for molecular exploration and optimization.
Recent studies incorporate experimental validation of ML-designed molecules and target the development of automated experimental molecular design guided by ML, e.g., in~\cite{Koscher2023}.
Overall, \emph{molecular ML for the transformation towards accelerated molecular design} shows great promise~\cite{du2024machine}. 

Within ChemE, it is advantageous and desirable to integrate molecule and process design, cf.~\cite{Bardow.2010, Zhang.2016, burger2015hierarchical}. 
Specifically, finding optimal chemical species for a process, i.e., molecules such as working fluids, solvents, and products, should be considered as an integrated part of process design.
To achieve this, the molecular properties that are relevant for the process are considered as part of the design formulation.
Both the molecular and the process structure are considered as degrees of freedom in the design.
To date, molecular properties in process models are typically calculated with established thermodynamic property models, e.g., NRTL~\cite{Renon.1968} and PC-SAFT~\cite{gross2001perturbed}.
These established models provide very accurate predictions but are limited to molecules for which experimental data is available.
In contrast, ML methods enable predictions for molecules not included in model training, e.g., see~\cite{Winter.2022, Rittig_GNNgammaIL.2022}.
Predictive group contribution methods like UNIFAC show lower accuracy than modern ML approaches and their applicability is limited to molecules for which model parameters are readily available~\cite{Jirasek.2020}, whereas ML models trained on large data sets typically provide a wider applicability range~\cite{SanchezMedina.2023, Jirasek.2020, Rittig_GNNgammaIL.2022, SanchezMedina.2022}.
Further approaches based on quantum mechanics and statistical thermodynamics, such as COSMO-RS~\cite{Klamt.2010}, can predict a wide range of molecules and properties but (in some cases) have been outperformed by ML, e.g., for activity coefficients~\cite{SanchezMedina.2023, Rittig_GNNgammaIL.2022} and solvation free energies~\cite{Vermeire.2021, leenhouts2025pooling}.
However, \emph{process modeling currently lacks state-of-the-art molecular ML models like GNNs}, as such models have not yet been integrated into process simulation software.
The identification of suitable chemical species for processes is therefore typically restricted to screening a list of known molecules with readily available property values or thermodynamic model parameters, not making use of recent developments in molecular design with ML.
We anticipate that \emph{the integration of ML for molecular property prediction and design with process design and optimization bears large potential and will advance chemical process engineering}. 

In the following, we discuss recent concepts in molecular ML and present a perspective on research directions to advance modeling and design at the molecular and process scale in ChemE.

\section{Machine Learning for Molecules and Mixtures}\label{sec:molecular_ML}

\noindent We first provide an overview of molecular ML methods for predicting properties of molecules and mixtures. 

\subsection*{Pure Species: From Structures to Properties}\label{sec:pure_comp}

\begin{figure*}[b!]
	\begin{center}
		\includegraphics[trim={0cm 2.5cm 0cm 3.5cm},clip, width=1\textwidth, keepaspectratio]{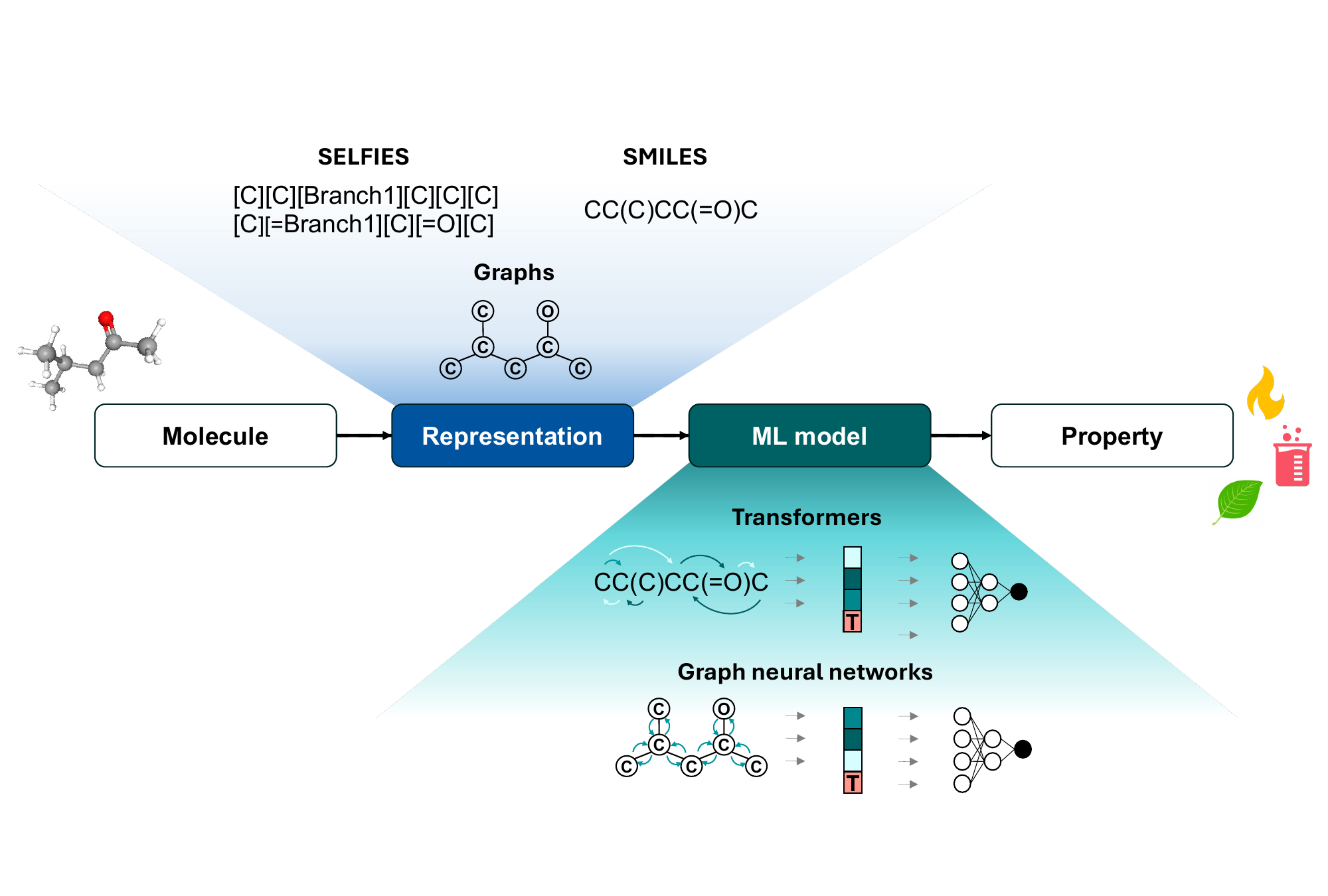}
		\caption{Schematic illustration of molecular machine learning approaches for property prediction: The molecule is represented in a machine-readable format, e.g., strings (here SELFIES~\cite{krenn2020self} and SMILES~\cite{weininger1988smiles}) or graphs, and then mapped to the property of interest by an ML model, e.g., a transformer or a graph neural network.}
		\label{fig:mol_to_prop}
	\end{center}
\end{figure*}

\noindent ML models enable to learn physicochemical properties directly from molecular structures of pure species, cf. overviews in~\cite{Reiser.2022, Rittig_GNNBook.2022}.
We show prominent end-to-end ML approaches for molecular property prediction in Figure~\ref{fig:mol_to_prop}.
The general idea of these end-to-end approaches is first to represent molecules in a machine-readable format.
The ML model then encodes this molecular representation into a continuous vector -- sometimes referred to as latent vector or learned molecular fingerprint -- in a learnable molecule-to-vector fashion, cf.~\cite{grohe2020word2vec}.
The learned vector can be combined with some state information, such as the temperature and pressure, e.g., by simple concatenation or a trunk network, cf.~\cite{Pavsek2025_DeepGraphNet}. 
The resulting vector is then mapped to the property of interest, typically by a standard neural network. %
As the ML model is trained in an end-to-end manner from structure to property~\cite{Rittig_GNNBook.2022}, the vector representation ideally captures the structural information relevant to the property of interest~\cite{Coley.2017}.
The structure-to-property characteristic is the key difference to traditional molecular ML based on molecular descriptors and static fingerprint approaches, like extended connectivity fingerprints (ECFPs)~\cite{rogers2010extended}.
While such traditional molecular ML approaches have a fixed way to transform the structure to a vector representation, the learnable molecule-to-vector encoding in modern ML models is more flexible and allows to achieve state-of-the-art accuracies -- if sufficient data is available for training.

In general, two main aspects characterize the type of ML approach: (i) the molecular representation and (ii) the model architecture for the learnable molecular encoding.

\textbf{(i) Molecular representation:} Commonly used representations for small molecules are based on strings, e.g., SMILES~\cite{weininger1988smiles} and SELFIES~\cite{krenn2020self}, and on geometry, e.g., molecular graphs and point clouds; for detailed overviews, we refer to~\cite{krenn2020self, wigh2022review, atz2021geometric, duval2023hitchhiker, alampara2025general}.
Representing molecules for ChemE applications poses a particular challenge when it comes to larger, more complicated structures, such as polymers and catalysts, as well as multiple interacting structures, as in mixtures and chemical reactions.
As such, adaptations of string- and graph-based representations are being actively developed, e.g., accounting for the stochastic nature of polymers~\cite{aldeghi2022graph, lin2019bigsmiles}. 

The used representation determines the level of prior structural information that is provided to the model:
Whereas string-based representations and molecular graphs capture the topological structure of molecules and can be enriched by additional descriptors, e.g., on stereochemistry, the full spatial information, i.e., the arrangement of atoms in 3D space, can be captured in point clouds and geometric graphs, which is highly useful or even necessary for accurate predictions of certain properties, such as electronic ones.
However, 3D information is typically not available and thus needs to be calculated with computationally costly quantum mechanical methods.
To speed up these calculations, quantum computing and machine learning interatomic potential (MLIP) approaches are promising and actively researched, cf. overviews in~\cite{cao2019quantum, anstine2025aimnet2, jacobs2025practical}.
Moreover, representing multiple interacting structures, e.g., in mixtures and reactions, in 3D comes with challenges, such as relative positions and orientation, and is thus also a current area of research, see, e.g.,~\cite{van20243dreact}.
Herein, we rather focus on string- and graph-based molecular representations, as predictions based on topological structural information have shown high accuracy for many properties relevant to ChemE.

\textbf{(ii) ML architecture:} The two most predominant deep-learning architectures for molecular property predictions are transformer architectures~\cite{vaswani2017attention, Winter.2022}, which originated in natural language processing, and geometric models, mainly graph neural networks (GNNs)~\cite{Heid.2024, Coley.2017, Gilmer.2017}, that respect the natural invariances of graph and spatial representations of the molecules.
In the learnable molecular encoding, structural information from the molecular representation is extracted:
GNNs extract structural information by passing information along edges in the molecular graph, where edges typically corresponding to chemical bonds and the process is referred to as message passing~\cite{Reiser.2022, Rittig_GNNBook.2022, Gilmer.2017}.
That is, GNNs assume that properties are primarily influenced by the set of local atom environments within a molecule. 
Thus, they come with a strong locality bias, similar to group contribution methods, but with a more flexible, self-evolving character.
In contrast, transformers explicitly consider both local and long-range interactions between atoms by the attention mechanism.
This also requires inferring chemical principles of bonds and locality from the training data, the chemical ``grammar''~\cite{alampara2025general}, typically resulting in the need for pretraining and higher data demands. 

Notably, from a methodological viewpoint, transformers can be considered as GNNs operating on fully connected graphs and using the attention-based message passing~\cite{joshi2025transformers}.
The two differences in molecular applications between GNNs and transformers lie in (1) the positional encoding typically used in transformers, which can break structural invariances of molecules (e.g., the same molecule can be represented with different SMILES, which lead to different prediction by transformers due to the positional encoding of atoms/tokens), and (2) the inductive bias, i.e., whether a molecular property is rather influenced by local atom environments (GNNs) or also by long-range atomic interactions (transformers).
Whether transformers are superior to GNNs in exploiting such long-range interactions is actively researched, e.g., in~\cite{tonshoff2023did}, and should be further investigated in the molecular context.
In this regard, graph transformers that combine the concept of local message passing on graphs with capturing long-range interactions through the attention-mechanism are also promising for molecular applications~\cite{rong2020self, sypetkowski2024scalability, anselmi2024molecular} but so far less explored in ChemE.
As for many molecular properties, the relationship between structure and property is not fully understood, e.g., if rather local or long-range interatomic effects are relevant, it is advisable to compare the approaches in practical applications. \\

\noindent Molecular ML approaches, such as GNNs and transformers, have recently been extensively applied for prediction of pure-component properties relevant to ChemE.
This includes numerous molecular types, such as small organic molecules~\cite{dou2023machine}, polymers~\cite{ge2025machine}, and ionic liquids~\cite{song2023computer}, and a variety of properties, such as boiling~\cite{Rittig_GNNBook.2022, hoffmann2025grappa} and melting points~\cite{sivaraman2020machine}, vapor pressures~\cite{hoffmann2025grappa, Lansford.2023, santana2024puffin}, density~\cite{Winter.2023_PCSAFT}, critical micelle concentration~\cite{qin2021predicting, Brozos_gnnsurfactant.2024}, toxicity~\cite{seal2025machine}, and biodegradability~\cite{Rittig_GNNBook.2022}.
In fact, the prediction capabilities of these ML models have been shown to exceed well-established prediction models based on COSMO, group contributions, and descriptors -- in terms of both accuracy and applicability range, see, e.g.,~\cite{hoffmann2025grappa, Winter.2023_PCSAFT}, if sufficient data is available for training (typically at least a few hundred data points are needed, cf.~\cite{Schweidtmann_FuelGNN.2020}).
In particular, they enable generalization to novel, unseen components, i.e., components for which experimental data is not readily available, given some kind of structural similarity to the molecules used for training.
We see the generalization capabilities of ML as most promising for the identification of novel, more sustainable chemical species, see, e.g.,~\cite{Koscher2023, peng2022human, konig2024machine}.

\subsection*{Properties of Mixtures}\label{sec:mixtures}
\begin{figure*}[hpb]
	\begin{center}
		\includegraphics[trim={2cm 11cm 2cm 0cm},clip, width=1\textwidth, keepaspectratio]{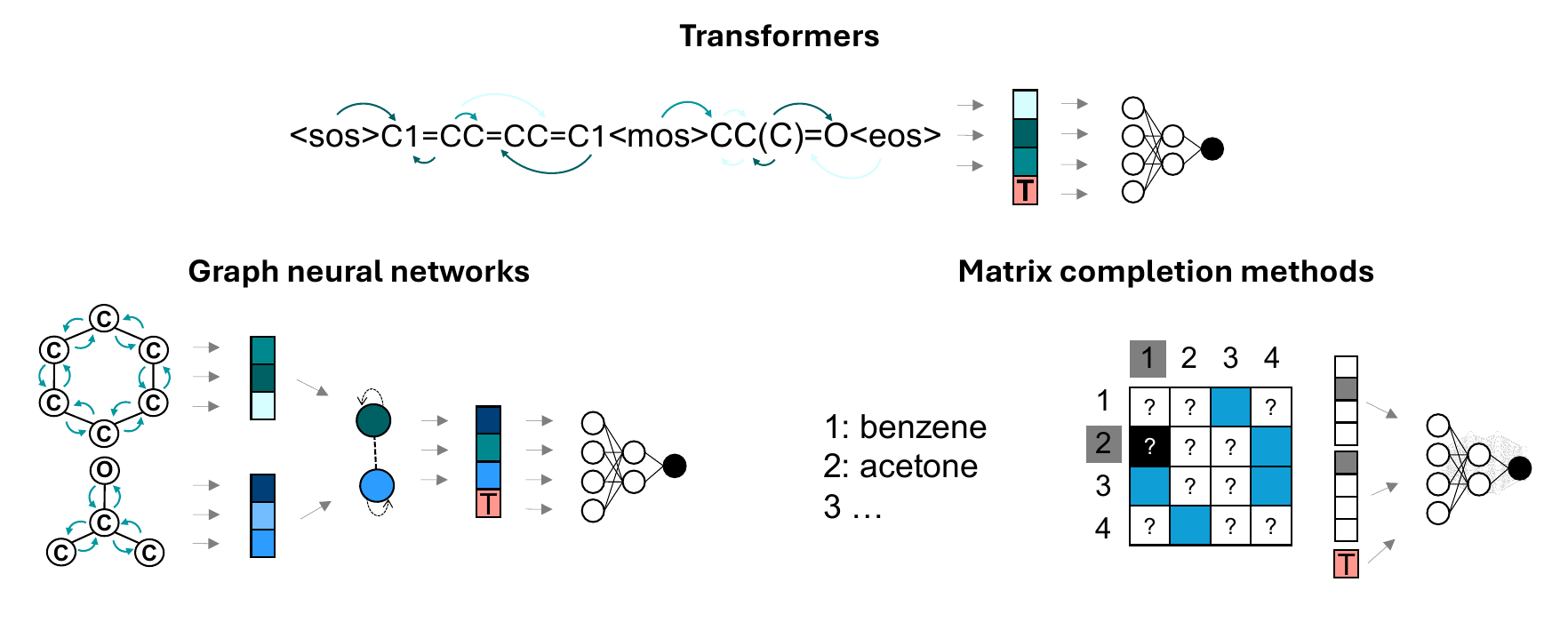}
		\caption{Schematic illustration of molecular machine learning approaches for predicting mixture properties at the example of a binary mixture of benzene and acetone.}
		\label{fig:mix}
	\end{center}
\end{figure*}

\noindent Molecular ML models have also been adapted to predict properties of mixtures, as illustrated in Figure~\ref{fig:mix}.
Here, next to GNNs and transformers, matrix completion methods (MCMs) have been used.
The ways to treat mixtures differ significantly between these methods.

In GNNs, the molecules within a mixture are first encoded to individual vector representations, analogously to pure-component property prediction.
The resulting molecular vectors are then aggregated to obtain a vector representing the mixture, the mixture fingerprint, which is then mapped to the mixture property.
Several ways of aggregating the molecular fingerprints have been proposed, e.g., concatenation or weighted sum of the molecular vectors, where the weights correspond to the molar fractions~\cite{leenhouts2025pooling, brozos2025predicting}.
Furthermore, mixtures themselves can also be represented by graphs, which allows to capture molecular interactions by applying GNNs before the aggregation step, cf.~\cite{SanchezMedina.2023, Qin.2023, Rittig_GibbsDuhemGNN.2023}.

Transformers treat mixtures as single instances, where the input is typically a sequence of SMILES of the molecules that contains special tokens indicating the start/end of a SMILES, see, e.g., Figure~\ref{fig:mix} and~\cite{Winter.2022}.
The model then applies the attention mechanism to the complete sequence, yielding a mixture vector.
Notably, transformers do not preserve the order invariance of mixtures, i.e., the same mixture can be represented in a different order (e.g., water/ethanol and ethanol/water), but a single sequence implies a fixed order.
This issue can be addressed by data augmentation, e.g., using multiple sequences for the same mixture with different orders during model training.
Here, it would be interesting to investigate architectural adaptions imitating the single molecule encoding and aggregation as in GNNs for mixtures.

MCMs, or more general tensor completion methods (TCMs), consider mixture property prediction as filling in the missing entries of a matrix, where the dimensions correspond to molecules and the entries to property values.
Notably, dimensions can also correspond to states such as temperature.
For the completion step, neural networks or Bayesian inference are often used cf.~\cite{Jirasek.2021, Chen.2021}.
In contrast to GNNs and transformers, MCMs typically do not consider any structural information of the molecules.
Rather, the molecules are simply represented as an index in the respective dimension of the matrix/tensor, which is analogous to a one-hot encoding. 
Thus, the applicability of MCMs is restricted to mixtures that are composed of molecules that occur in the training data set, hence, predicting properties of mixtures with unseen molecules, as with GNNs and transformers, is not possible. 
To address this issue, it would be interesting to replace the one-hot encoding in MCMs with molecular fingerprints in future work.

All three methods, GNNs, transformers, and MCMs, have been extensively applied to mixture property prediction.
In particular, the activity coefficient of binary mixtures has been targeted, at infinite dilution~\cite{Winter.2022, Jirasek.2020, SanchezMedina.2022}, varying temperature~\cite{SanchezMedina.2023, Rittig_GNNgammaIL.2022, Chen.2021}, varying composition~\cite{rittig2024thermodynamics, Qin.2023, Rittig_GibbsDuhemGNN.2023}, and all together~\cite{specht2024hanna, Winter.2023}, resulting in high accuracies of molecular ML, beyond COSMO-RS~\cite{Klamt.2010} and UNIFAC~\cite{Fredenslund.1975}.
Applications further include properties of mixtures with varying numbers of components, e.g., solvation free energies~\cite{leenhouts2025pooling} and critical micelle concentration~\cite{brozos2025predicting}.
As the combinatorial space of mixtures is vast, especially with an increasing number of components, the high accuracies of molecular ML models are highly promising to accelerate the search for mixtures with desired properties for ChemE applications. \\

\subsection*{Research directions}

\begin{figure*}[htpb]
	\begin{center}
		\includegraphics[trim={4cm 8.5cm 5cm 2cm},clip, width=1\textwidth, keepaspectratio]{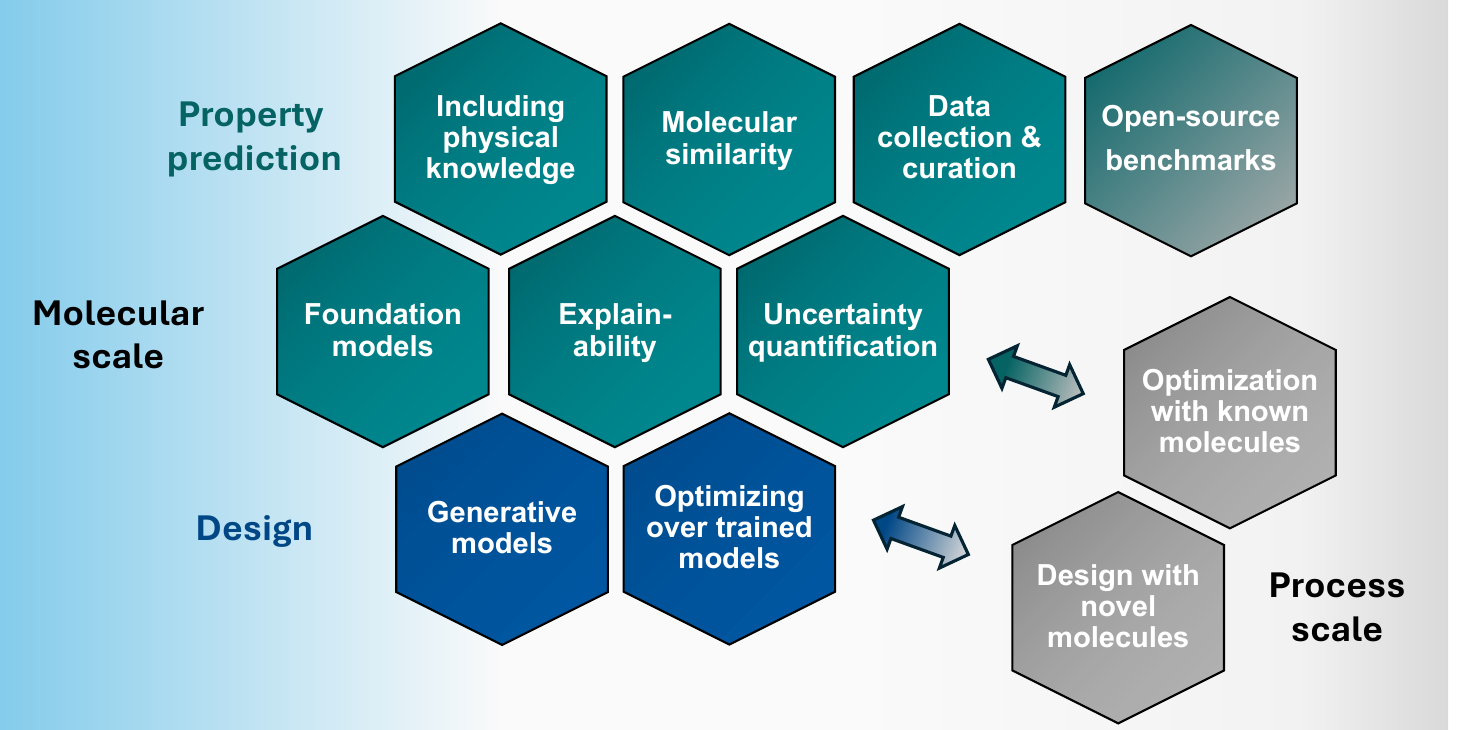}
		\caption{Overview of research areas and directions for molecular ML in chemical process engineering.}
		\label{fig:overview}
	\end{center}
\end{figure*}

\noindent With predictive molecular ML models having shown remarkable results, we see numerous research directions for further advancements and integration with the process scale, as indicated in Figure~\ref{fig:overview}.
In the following sections, we identify several research areas that are highly important for the transfer to practical usage and promise further advancements.
Furthermore, we argue that molecular ML can substantially contribute to two main objectives of ChemE: the design of more sustainable molecules with desired properties, and the integration into process design and optimization, which we respectively discuss in Sections~\ref{sec:mol_design}~\&~\ref{sec:proc_design}.  

\section{Advancing Predictive Models}\label{sec:advancing_molml}

\noindent We first present research areas that promise advancement in predictive molecular ML models: including physicochemical knowledge, data collection \& curation that enable benchmarks, foundation models, explainability, uncertainty quantification, and similarity. \\

\noindent \textbf{Including physicochemical knowledge} will be essential to further advance molecular ML.
In recent years, ML models have been adapted to account for physicochemical principles. 
For example, GNN architectures have been adapted to preserve physical symmetries of molecules, i.e., rotational and translational invariance~\cite{duval2023hitchhiker, rittig_PhDthesis}, account for stereochemical arrangements~\cite{adams2021learning}, and consider the influence of the molecular size~\cite{Schweidtmann.2023}.
Also structural characteristics of certain types of molecules, such as polymers and surfactants, have been used to refine ML architectures~\cite{aldeghi2022graph, brozos2025predicting}.
These architectural adaptations can have a significant influence on the consistency and accuracy of the property predictions and should therefore be further explored. 

\begin{figure*}[tpb]
	\begin{center}
		\includegraphics[trim={2cm 6cm 2cm 1cm},clip, width=1\textwidth, keepaspectratio]{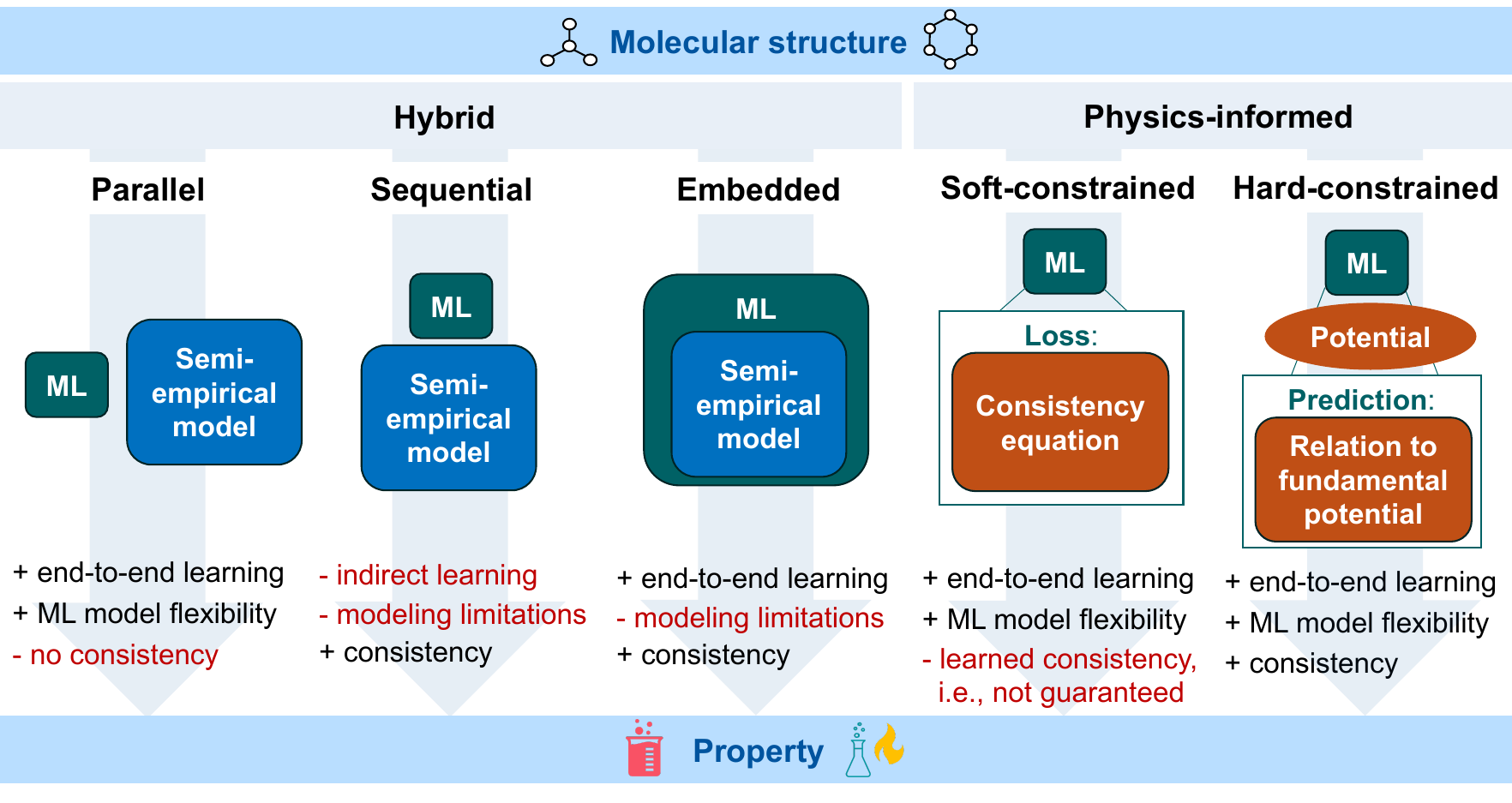}
		\caption{Overview of hybrid and physics-informed approaches to combine physicochemical knowledge with molecular machine learning models, along with respective advantages and disadvantages.}
		\label{fig:hybrid}
	\end{center}
\end{figure*}

Additionally, hybrid and physics-informed molecular ML approaches have been proposed, as we illustrate in Figure~\ref{fig:hybrid}, also cf. overview in~\cite{Jirasek.2023}.
\emph{Hybrid models} include the combination of a molecular ML with a semi-empirical model in either a sequential, parallel, or embedded way. 
In the parallel setting, the ML model predicts the error of a semi-empirical model, e.g., of COSMO-RS or UNIFAC, as in~\cite{SanchezMedina.2022}; the flexibility of ML is preserved but not constrained by any physical knowledge, hence predictions are likely to be physically inconsistent.
The sequential approach is characterized by predicting the parameters of a semi-empirical model, while the embedded approach incorporates semi-empirical equations into the ML architecture, such as NRTL or PC-SAFT equations as in~\cite{Winter.2023_PCSAFT, Winter.2023}.
Both approaches ensure coherence with the physical knowledge -- given that the semi-empirical model is physically consistent; however, the prediction accuracy is constrained by the limitations of the semi-empirical model.
We note that the embedded approach should usually be preferred over the sequential approach, as it can be directly trained on property data.

We see two promising research directions building on hybrid approaches: 
First, applying explainability methods to ML models in the parallel approach can elucidate information on error sources and lead to mechanistic insights.
Secondly, the sequential and embedded approach can be used to predict semi-empirical model parameters for molecules for which experimental data is missing and classical parameter fitting is not possible, cf.~\cite{Winter.2023_PCSAFT, Felton_MLSAFT.2023, Habicht2023}, so they can directly be utilized in process simulation software, which we explain in more detail in Section~\ref{sec:proc_design}.

In contrast to building on semi-empirical models, \emph{physics-informed ML} incorporates algebraic and/or differential relations to fundamental properties.
For example, the Helmholtz free energy is a thermodynamic potential from which related properties can be deduced by applying fundamental thermodynamics: intensive properties such as entropy or internal energy are related to first-order derivatives of the Helmholtz free energy; heat capacity and thermal expansion coefficient to second-order derivatives.
To incorporate such fundamental relations, two approaches have emerged: soft-constrained and hard-constrained ML.
Soft-constrained ML uses relations to fundamental properties as a regularization term in the loss function, i.e., the model learns to provide predictions that follow these relations -- similar to physics-informed neural networks (PINNs).
Hard-constrained ML describes the concept of embedding relations to fundamental relations into the ML architecture, e.g., in~\cite{Rosenberger.2022}, or through projection layers, e.g., in~\cite{lastrucci2025enforce, iftakher2025physics}, guaranteeing physical consistency.
Both approaches have recently been successfully applied for the prediction of thermodynamic properties of fluids and solids~\cite{Chaparro.2024} and activity coefficients in binary mixtures~\cite{rittig2024thermodynamics, specht2024hanna, Rittig_GibbsDuhemGNN.2023}.

Extending physics-informed molecular ML to further properties and accounting for state transitions~\cite{Chaparro.2024} will be critical for ensuring physical consistency and achieving higher prediction accuracies while decreasing data demands for training.
Physical consistency will also increase acceptance, safety, and trust in molecular ML applications for practitioners. \\

\noindent \textbf{Property data collection \& curation} is critical for advancing molecular ML. 
In our opinion, data scarcity remains the major limiting factor in advancing property prediction for ChemE applications.
In fact, sophisticated ML approaches are readily available for molecular applications, but a large fraction of the experimental molecular and mixture property data relevant for ChemE is distributed in numerous literature sources, commercial datasets, and private datasets of chemical companies. 
Obtaining additional experimental data is naturally costly and labor-intensive, so it is required to increase efforts in leveraging existing data.
For publicly available data in the literature, we see large potential of automatic extraction by agentic ML frameworks~\cite{ramos2025review}; yet, human intervention and curation will be required, as reported experimental data can have errors, e.g., caused by unit conversion.
We advocate for efforts from both academia and industry to assemble such data and make it available for the development of prediction models.
For this, the recent collection of data relevant to the chemical science, the ChemPile data set~\cite{mirza2025chempile}, can serve as a blueprint.
To additionally utilize proprietary property data, federated learning projects should be initiated that enable and incentivize databank organizations and chemical companies to contribute to the development of ML models without sharing their sensitive data, cf.~\cite{dutta2025federated, rittig2025federated}.
By scaling the amount of high-quality data that can be utilized for training, the accuracy and applicability domains of ML models will increase.\\

\noindent \textbf{Benchmarks} will catalyze the development of molecular ML models for ChemE applications.
Well-defined benchmarks will motivate both the ChemE and the ML community to further develop molecular ML models and reach state-of-the-art accuracies, as can be seen at the prime example of the QM9 dataset~\cite{Ramakrishnan2014, Ruddigkeit2012}.
In addition, comparing currently available molecular ML models is difficult, as many ChemE-related property data sets used for training and testing are not provided as open-source.
We advocate that property prediction benchmarks are created in collaboration with the chemical industry to also include requirements for industrial applications.
These benchmarks should cover a wide spectrum of molecule classes and properties relevant for ChemE.
Specifically, benchmarks should not only focus on prediction accuracy but also account for physical/thermodynamic consistency.
They should further provide multiple test sets to investigate different scenarios, i.e., interpolation/extrapolation of state variables, generalization to novel, unseen molecules, etc.
Open-source property prediction benchmarks can thereby direct the development of new molecular ML approaches to meet industrial needs and criteria for practical applications. \\

\noindent \textbf{Foundation models} promise to advance molecular ML by training on large amounts of property data, so that they can generalize and be fine-tuned on specific property prediction tasks, even if only little data is available~\cite{rishi2021foundation}.
To date, the typical approach in molecular ML is to train predictive models from scratch, i.e., for a property prediction task at hand, data is collected and then an ML model is trained, often using readily implemented molecular ML frameworks like chemprop~\cite{Heid.2024, yang2019analyzing}.
This can be challenging as property data is often scarce in ChemE applications.
To address this issue, mainly self-supervised, transfer, multi-task, and multi-fidelity learning approaches have been investigated in the molecular domain, e.g., in~\cite{Vermeire.2021, rong2020self, chithrananda2020chemberta, burns2025descriptor}.
While in some cases, these methods can lead to improvements in the accuracy and applicability range of the prediction, the experimental data used is rather small and covers only a few properties, which limits generalization.
Further cases show that combining property data as in multi-task models can also decrease model performance, as the optimization during training becomes more difficult, see e.g.,~\cite{Heid.2024, bin2025exploring}.
We argue that using additional well-curated data should generally increase -- at least not harm -- model performance, so further developments in scaling molecular ML architectures and improving training procedures, e.g., recently proposed task-specific early stopping~\cite{eraqi2025molecular}, are needed to utilize the information and relationships hidden in molecular property data sets.

Recently, a few studies have assembled large molecular data sets and trained transformers and GNNs on molecular property prediction, e.g.,~\cite{sypetkowski2024scalability, beaini2023towards, klaser2024texttt}. 
However, these studies mostly focus on properties relevant for biological and pharmaceutical applications, whereas chemical engineering lacks large data sets.
Thus, we believe that assembling large data sets with properties relevant for chemical engineering by combining data for different properties bears large potential for ML. 
In fact, we hypothesize that increased diversity of molecular classes and properties could enable ML to exploit and uncover chemical patterns beyond simple property correlations, which will facilitate generalization; whether this will be in the sense of a foundation model remains to be explored.
In particular, we see great potential in including \emph{fundamental properties}, such as Gibbs and Helmholtz free energies, as these provide insights on relations between different properties and enable consistent predictions~\cite{rittig_PhDthesis}.
Overall, combining molecular machine learning (ML) models with fundamental property relations and training them on large-scale property datasets can advance property prediction in small data applications and increase generalization capabilities. \\

\noindent \textbf{Explainability} promises to uncover unknown molecular structure-property relationships.
Explaining and interpreting predictions of molecular ML models has been actively researched in recent years, e.g., by investigating model sensitivities through gradient-based or counterfactual methods, cf.~\cite{RodriguezPerez.2021, Wellawatte.2023}.
So far, most approaches focus on explaining local, single-instance predictions, i.e., for a given molecule, which is useful in validating whether an ML model correctly identifies structural parts of the molecule that are known to influence a property.
We advocate to also focus on systematically explaining ML predictions on a global model level~\cite{Yuan.2023}, that is, finding generalizable structure-property relations that hold for a diverse collection of molecules.
For example, subgraphs representing molecular motifs can be clustered and analyzed with large language models as in~\cite{teufel2024global}, or learned molecular vectors can be utilized in hierarchical clustering to identify property-specific molecular classes~\cite{gond2025hierarchical}.
Such research should also distinguish between abductive (``Which structural parts support a certain property prediction?'') and contrastive relations (``Which structural parts need to be changed to get a different property prediction?'')~\cite{ignatiev2020contrastive, Wellawatte.2022}.

A significant aspect to consider here is that ML models are highly overparameterized and capture nonlinearities in large data sets that enable reaching accuracies beyond mechanistic models.
So, we hypothesize that related phenomena are difficult to translate to high-level explanations for humans.
Important questions that should be addressed here are: 
What are the explanations for the accuracy gains beyond mechanistic models?
To what extent can we achieve a mechanistic understanding with ML models, and are explainability and high accuracy conflicting objectives at some point? \\

\noindent \textbf{Uncertainty quantification} is highly important for practical applications of molecular ML.
Numerous uncertainty quantification methods have been investigated for molecular ML, including similarity- and ensemble-based methods, mean-variance estimation, and conformal prediction, with ambiguous results regarding superiority and usability, cf. overview in~\cite{Hirschfeld.2020}.
The field continues to be actively researched, and promising methods are being proposed frequently for ChemE applications, e.g., based on architecture search~\cite{jiang2024uncertainty} or stochastic gradient Hamiltonian Monte Carlo~\cite{gao2025uncertainty}.
A major challenge here is to decompose the uncertainty in the epistemic part caused by the model and the aleatoric part~\cite{Heid.2023}, i.e., the uncertainty inherent to the property data due to different experimental setups, instruments, reporting, etc.
To elucidate model uncertainties, quantification methods have very recently been coupled with explainability approaches, see, e.g.,~\cite{komissarov2025explainable, teufel2025improving}.

It will be particularly interesting to test developed uncertainty quantification methods on property prediction benchmarks created for ChemE. 
We also stress that it is promising to test and further develop these methods for multi-task models, and ultimately foundation models, since this could help to infer uncertainty relationships between different properties and facilitate identifying erroneous data points. \\

\noindent \textbf{Molecular similarity} based on molecular fingerprint vectors can reveal novel chemical relations that are learned by molecular ML models.
The concept of similarity is frequently used in analyzing molecular latent spaces.
Specifically, the learned fingerprint vectors in molecular ML models are reduced to a few (typically two) dimensions and visualized in a human-understandable way.
Then, distances between individual vectors can be interpreted as molecular similarity specific to the physicochemical property the molecular ML model is trained on, potentially revealing chemical insights, e.g., clusters of molecular classes~\cite{Vermeire.2021}.
Furthermore, similarity-based approaches can be used to assess model uncertainties; that is, predictions are assumed to have higher accuracy for molecules that are encoded into a fingerprint vector close to those of the molecules used for training, cf.~\cite{Hirschfeld.2020}.
Analyzing molecular similarity is thus highly related to explainability and uncertainty quantification research.

We see further need in investigating molecular similarity in learned fingerprint/latent spaces, as the learnable molecule-to-vector encoding is an essential part of molecular ML models -- distinguishing them from established, static fingerprint approaches.
The following specific questions should be addressed:
What are the differences in the distance of the learned fingerprint vectors depending on the property to be predicted?
Is the fingerprint vector space actually interpretable, given its typical high dimensionality, and how does the number of dimensions influence the similarity?
How does the learned property-oriented similarity relate to structural similarity based on static fingerprints, as well as to more abstract concepts, such as string or graph similarity?
Lastly, it would be interesting to research similarity in multi-task models.
For example, do shared model layers capture high-level chemical concepts that are relevant to subsets or all of the considered properties, and how do the fingerprint vectors change in property-specific layers? 
Addressing these questions would greatly increase the understanding and interpretability of molecular ML models. \\

\noindent Overall, we find many promising research directions that have the potential to increase predictive capabilities of molecular ML and uncover novel chemical insights.
In particular, we see the need for strong collaboration between academia and industry to improve the development and reliability of molecular ML models, which will eventually lead to their practical application on an industrial scale.

\section{Designing Molecules with Desired Properties}\label{sec:mol_design}

\noindent While molecular ML models enable property prediction, it is desirable to identify molecules with optimal properties for specific ChemE applications, referred to as CAMD.
For this, we identify two actively investigated approaches: (i) generative ML models and (ii) deterministic global optimization over molecular ML models. \\

\noindent \textbf{Generative ML} models propose new molecules with desired properties by learning from existing molecular structures.
Notably, they can explore the chemical space beyond established CAMD approaches in ChemE, e.g., based on structure enumeration or group contribution models embedded into optimization formulations, which are restricted to a combination of functional groups~\cite{mann2023group}.
Over the last years, numerous generative molecular ML models have been proposed, including variational autoencoders (VAEs), reinforcement learning (RL), generative adversarial networks (GANs), and recently diffusion- and flow-based models, cf. overviews in~\cite{Bilodeau2022, du2024machine}. 

In ChemE, generative models have recently been applied for identifying promising molecules as fuels~\cite{RittigRitzert_GraphMLFuel.2022, sarathy2024artificial}, polymers~\cite{vogel2025inverse}, and solvent~\cite{pirnay2025graphxform}.
As generative ML models can also account for synthesizability or constraints on molecular motifs/building blocks~\cite{pirnay2025graphxform, guo2025generative, tu2025askcos}, they are particularly promising to accelerate molecular discovery.
Future works in ChemE should target experimental validation, as in~\cite{Koscher2023, RittigRitzert_GraphMLFuel.2022}. \\

\noindent \textbf{Global optimization with molecular ML models embedded} enables finding molecules with globally optimal properties.
Specifically, ML models that are trained to predict molecular properties can be embedded into optimization formulations for molecular design.
As such, the ML model weights are fixed, and the prediction is optimized as a function of the inputs, i.e., the molecular structure is the degree of freedom; notably, the prediction can also be considered as a constraint in a design formulation.
For example, trained GNNs have been embedded into molecular design formulations, which enables to find global optimal molecules as predicted by the GNN using deterministic solvers~\cite{McDonald_OptGNN.2024, Zhang_OptGNN.2023}.
For this, two major challenges arise: additional constraints need to be formulated to restrict the search space of molecular structures to chemically valid molecular graphs, and the highly nonlinear GNN layers are part of the problem formulation, making solving computationally costly and currently impractical for molecules with more than a few atoms~\cite{McDonald_OptGNN.2024, Zhang_OptGNN.2023, rittig2024deterministic}. 
As an alternative approach that circumvents these challenges, VAEs for molecule generation can jointly be trained with neural networks for property prediction on the VAE's latent space, which allows to only consider the neural network in the molecular design formulation, cf.~\cite{rittig2024deterministic, wang2022novel}.
This approach, however, comes with the additional computational costs and difficulty of training a VAE jointly with a neural network.

Overall, optimizing over molecular ML models is highly promising for molecular design, as it enables finding optimal molecules (as predicted).
We advocate for further research in this area, including the embedding of other molecular ML models into optimization formulations, such as transformers and MCMs.
Embedding molecular ML models into optimization formulations will also be of major importance in process design, which we will discuss next.

\section{Towards Integration with the Process Scale}\label{sec:proc_design}

\noindent The integration of molecular ML into the process scale of ChemE is still in its infancy.
In fact, molecular ML is rarely used for: (i) predicting the properties of chemical species used in chemical processes in the context of process modeling and optimization; and
(ii) designing molecules and mixtures as an integrated part of process design, known as computer-aided molecular and process design (CAMPD), cf. overviews in~\cite{scheffczyk2017cosmo, adjiman2025challenges, iftakher2023overview, rehner2023molecule, bosetti2025integrated}.
In fact, CAMPD approaches typically embed molecular fragmentation and group contribution methods in optimization formulations, limiting the molecular and process design space.

We advocate for the integration of molecular ML into process models, e.g., for modeling thermodynamic properties and sustainability factors.
Here, we distinguish two scenarios, considering (i) known molecules in practical use, and (ii) generalizing to novel molecules. \\

\noindent \textbf{Known molecules in practical use} within chemical processes typically come with readily available property data.
This means that established semi-empirical models, e.g., based on equation-of-state approaches, or surrogate models, such as polynomials and shallow neural networks, can be fitted to this data.
These models typically provide reasonably accurate property predictions for process optimization and design, without the need for advanced molecular ML approaches.

However, in cases where the property data or semi-empirical/surrogate models are limited to specific state ranges, e.g., in terms of temperature and pressure, process optimization becomes restricted.
Molecular ML models trained on a diverse set of molecules with corresponding properties can provide state-dependent predictions for wider ranges than a model fitted only on property data of the single molecule (or mixture) of interest.
As such, molecular ML models need to be embedded into process model formulations, e.g., using tools such as OMLT~\cite{ceccon2022omlt} or MeLOn~\cite{melon}, which will require model adjustments and might cause additional computational costs, cf.~ e.g.,~\cite{McDonald_OptGNN.2024, Zhang_OptGNN.2023}.
Alternatively, semi-empirical model parameters for the molecule (or mixture) of interest can be fitted to predictions of molecular ML models or extracted from hybrid molecular ML architectures that are trained on a more diverse set of molecules and wider state ranges, cf.~\cite{Winter.2023_PCSAFT, Felton_MLSAFT.2023} and Section~\ref{sec:molecular_ML}.
These parameters can then be directly used in process modeling without any additional model adjustments.
Therefore, molecular ML can expand process optimization and design to include wider operating ranges, potentially leading to increased process efficiencies. \\

\noindent \textbf{Generalizing to novel molecules} with desired properties for chemical processes is highly desirable.
For this, process performance indicators can be included in molecular design objectives (and constraints), e.g., solubility or partition coefficients for finding suitable solvents in separation processes, cf.~\cite{konig2024machine, pirnay2025graphxform}.
Further properties, e.g., accounting for the environmental impact of the proposed molecules, can also be included in the design.
Predictive and generative molecular ML models can then be employed and accelerate the identification of novel, sustainable solvents, reactants, catalysts, etc. that optimize these process performance indicators, whereas experimental validation remains critical.

Ultimately, we anticipate that ML-driven CAMPD will play an important role in ChemE, i.e., molecules and processes are designed simultaneously through ML.
In addition to embedding molecular ML models into process models, similarly to the approach for known molecules described above, the molecular structure becomes a degree of freedom, making optimization much more challenging and thus requiring further research.
To circumvent embedding equations of molecular ML models into optimization formulations, sequential ML-CAMPD workflows can be employed, as very recently proposed in~\cite{bosetti2025integrated} for the design of solvent-antisolvent mixtures and crystallization processes.
Specifically, molecular ML can be utilized in an iterative manner by (1) proposing molecular structures by molecular design algorithms, (2) predicting the properties of the proposed structures by predictive ML models, and (3) solving a process design formulation using these predicted properties, which then serves as a feedback for the design algorithm in (1)~\cite{bosetti2025integrated}, thereby integrating process design goals into molecular design.

Another particularly promising direction is to couple molecular ML with recently proposed generative ML approaches for process design~\cite{schweidtmann2024generative}.
Such ML-based process design approaches, mostly based on reinforcement learning, so far only include process variables as part of the design space~\cite{Gao.2024, Gottl.2023}.
Developing generative ML methods that enable the simultaneous design of molecules and processes bears large potential in automating and advancing CAMPD.

\section{Concluding Remarks}\label{sec:Conclusion}

\noindent ML has advanced molecular property prediction and design in ChemE by learning from data on molecules and mixtures.
We hypothesize that there are more chemical relationships hidden in these data than current molecular ML models have learned.
Data collection and curation is needed so that ML models can leverage and reveal these relationships through advanced model architectures that are based on physicochemical knowledge and model-level explainability.
Furthermore, coupling molecular ML with the process scale will accelerate the identification of novel, more sustainable molecules and mixtures that also lead to more efficient processes. 
It is of major importance that academia closely collaborates with the chemical industry to further advance molecular ML models and establish benchmarks for practical application in process design and optimization. 

Future work should focus on integrating generative ML for the molecular and process scale.
To this end, we see great potential in multi-agent frameworks~\cite{alampara2025general, rupprecht2025multi} for orchestrating and automating ChemE design tasks.

\section*{Declaration of Competing Interest}
\noindent The authors declare that they have no known competing financial interests or personal relationships that could have appeared to influence the work reported in this paper.

\section*{Acknowledgments}
\noindent This project was funded by the Deutsche Forschungsgemeinschaft (DFG, German Research Foundation) – 466417970 – within the Priority Programme ``SPP 2331: Machine Learning in Chemical Engineering''. \\
This work was also performed as part of the Helmholtz School for Data Science in Life, Earth and Energy (HDS-LEE).\\
The project was also funded by the European Union (ERC, SymSim, 101054974). 
This work was further funded by the Swiss Confederation under State Secretariat for Education, Research and Innovation SERI, participating in the European Union Horizon Europe project ILIMITED (101192964). 
Views and opinions expressed are however those of the author(s) only and do not necessarily reflect those of the European Union or the European Research Council. Neither the European Union nor the granting authority can be held responsible for them. \\
Funding by the Werner Siemens Foundation within the WSS project of the century ``catalaix'' is acknowledged. \\ 
MD and AM received funding from the Helmholtz Association of German Research Centers. \\
PS acknowledges support from the NCCR Catalysis (grant number 225147), a National Centre of Competence in Research funded by the Swiss National Science Foundation. \\

\bibliographystyle{unsrt}  
\bibliography{references}  

\end{document}